\documentclass[twocolumn,showpacs,preprintnumbers,amsmath,amssymb,prl]{revtex4}

\usepackage{graphicx}
\usepackage{bm}
\usepackage{overpic}
\usepackage{upgreek}
\usepackage{color}

\def\k{{\bf k}}
\def\q{{\bf q}}
\newcommand{\ua}{\uparrow}
\newcommand{\da}{\downarrow}

\begin{document}

\title{
\bf\Large{Fractional Flux Quantization in Loops of Unconventional Superconductors}}

\author{F. Loder$^{1,2}$}
\author{A. P. Kampf$^2$}
\author{T. Kopp$^1$}
\affiliation{
$^1$Experimental Physics VI, $^2$Theoretical Physics III,\\Center for Electronic Correlations and Magnetism, Institute of Physics, University of 
Augsburg, 86135 Augsburg, Germany}

\date{\today}

\begin{abstract}
The magnetic flux threading a conventional superconducting ring is typically quantized in units of $\Phi_0=hc/2e$. The factor 2 in the denominator of $\Phi_0$ originates from the existence of two different types of pairing states with minima of the free energy at even and odd multiples of $\Phi_0$. Here we show that spatially modulated pairing states exist with energy minima at fractional flux values, in particular at multiples of $\Phi_0/2$. In such states  condensates with different center-of-mass momenta of the Cooper pairs coexist. The proposed mechanism for fractional flux quantization is discussed in the context of cuprate superconductors, where $hc/4e$ flux periodicities as well as uniaxially modulated superconducting states were observed.
\end{abstract}


\maketitle

In superconducting (SC) rings flux quantization in units of $\Phi_0=hc/2e$ is usually attributed to the charge $2e$ of Cooper pairs carrying the supercurrent. This connection has been anticipated by F. London~\cite{London} long before the experimental confirmation of the SC flux quantum $\Phi_0$ by Doll and N\"abauer~\cite{Doll} and by Deaver and Fairbank~\cite{Deaver} in 1961. In the same year, $\Phi_0$ was derived from the BCS pairing theory as a consequence of the existence of two classes of SC wave functions with energy minima at even or odd multiples of $\Phi_0$~\cite{Byers,onsager:61,brenig:61}.

\begin{figure}[b]	
\centering
\includegraphics[width=0.65\columnwidth]{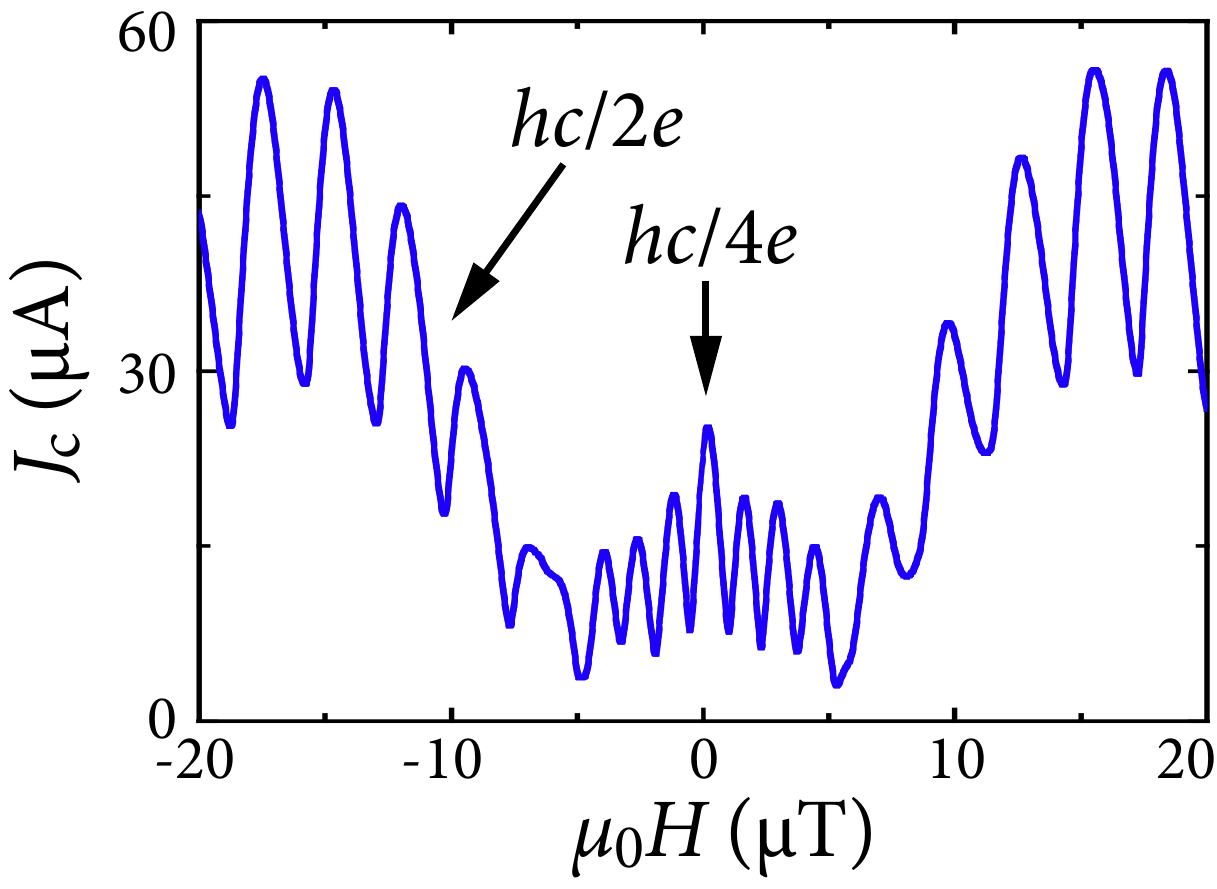}
\vspace{-1mm}
\caption{Critical current $J_{\rm c}(H)$ through a 24$^\circ$ YBCO grain-boundary SQUID at $T=4.2\,$K from Ref.~[\onlinecite{schneider:04}]. Clearly visible is the abrupt change of periodicity at $\mu_0H\approx\pm5\,\upmu$T. For the SQUID size in this experiment, a flux of $\Phi_0$ is achieved for a magnetic field of $2.7\,\upmu {\rm T}$. (Courtesy of C. Schneider)
\vspace{-3mm}}
\label{fig1}
\end{figure}

The formation of a pair condensate alone does not inevitably imply that $\Phi_0$ is the unit of flux quantization. In fact, this conclusion is only valid for a uniform condensate of non-interacting Cooper pairs, as assumed in BCS theory.  Normal persistent currents can as well sustain a $hc/2e$ flux periodicity in special geometries~\cite{riedel:93} or due to electron-electron interactions~\cite{fye:91,ambegaokar:91}. Correlations between Cooper pairs may also lead to fractional flux quanta. Such effects were first discussed by Little in 1964 in the context of fractional flux periodicities in the critical temperature of conventional SC cylinders~\cite{little:64}. More recently, unusual flux periodicities were recently reported in SQUID experiments with high-$T_{\rm c}$ cuprates~\cite{lind:03,schneider:04,schneider:05}. An example for $\Phi_0/2$ oscillations of the critical current is shown Fig.~\ref{fig1}. The low magnetic field data indicate a $\sin 2\varphi$ Josephson relation. As one likely origin of the fractional periodicity, multiple Andreev scattering in grain-boundary Josephson junctions was proposed~\cite{lind:03,schneider:04}; however, the abrupt disappearance of the $\Phi_0/2$ periodicity beyond a threshold field has remained unexplained.
In the search for an alternative origin, interactions between Cooper pairs and quartet formation have been investigated theoretically~\cite{koh:92,mackowiak:99,tarasevicz:04.2,aligia}. Yet, their influence on flux quantization is unresolved.

Here we formulate a conceptually distinct mechanism for fractional flux quanta in superconductors with a spatially modulated SC order parameter (OP), in particular for $\Phi_0/2$ flux periodicity and a $\sin 2\varphi$ Josephson relation in SQUIDs. 
The proposed concept rests on the coexistence of pair condensates with different center-of-mass momenta (COMM) of Cooper pairs and is unique for superconductors with unconventional pairing symmetries.

Consider the uniform OP $\Delta(\theta)=\Delta e^{i\theta(q-\Phi/\Phi_0)}$ for electron pairs with quantized COMM $\hbar q$ and angular coordinate $\theta$ on a hollow cylinder threaded by the magnetic flux $\Phi$. In this geometry, $q$ refers to the phase winding number of the OP upon circulating the cylinder once. The free energy $F$ of this system has a minimum for a vanishing phase gradient, i.e., at the flux value $\Phi=q\Phi_0$. Since $q$ is an integer, the fluxoid threading the cylinder is quantized in units of $\Phi_0$~\cite{schrieffer,degennes}. 

If instead the magnitude $\Delta$ of the OP oscillates around the cylinder, Cooper pairs with different COMMs exist. An example for a sign changing OP is
\begin{multline}
\tilde\Delta(\theta)=\left(\Delta_1e^{iq_1\theta}+\Delta_2e^{iq_2\theta}\right)e^{-i\theta\Phi/\Phi_0}\\
\xrightarrow{\Delta_1=\Delta_2}2\Delta_1\cos\left(\theta[q_1-q_2]/2\right)e^{i\theta([q_1+q_2]/2-\Phi/\Phi_0)}.
\label{intr1}
\end{multline}
In such a state, pairs with COMM $q_1$ coexist with pairs with $q_2$; its phase-winding number is $[q_1+q_2]/2$ which implies that $F$ has a minimum at $\Phi/\Phi_0=[q_1+q_2]/2$. Since $q_1+q_2$ can be both even or odd, the flux is hence quantized in units of $\Phi_0/2$.

The above state generalizes the time-reversal symmetric ``pair-density wave'' (PDW) concept with $q_1=-q_2$ (see e.g. Refs.~\cite{agterberg:08,berg:09} and references therein). This state bares similarities to Fulde-Ferrell-Larkin-Ovchinnikov states in an external magnetic field~\cite{fulde:64,larkin:64}. Due to its concomitant charge-density modulations, the PDW was proposed as a candidate groundstate of striped cuprate superconductors~\cite{,berg:07,berg:09}. For unconventional pairing symmetries with gap nodes~\cite{loder:10} it was indeed verified that the PDW state has a finite range of stability. A Ginzburg-Landau analysis furthermore demonstrated that the PDW state can sustain a $\Phi_0/2$ vortex phase~\cite{agterberg:08,agterberg:09}. The melting of the PDW state may even give rise to a state with charge-$4e$ superconductivity, which also leads to a $\Phi_0/2$ periodicity in a SQUID geometry~\cite{berg:09.2}.

Another system where the generalized PDW concept may lead to $\Phi_0/2$ flux quanta is the spin-triplet superconductor Sr$_2$RuO$_4$. Here it is the two-component order parameter $\{\Delta_{q_1}^{\ua\ua},\Delta_{q_2}^{\da\da}\}$ for the $s_z=1$ and the $s_z=-1$ condensates which allows for the coexistence of different COMMs~\cite{vakaryuk:11}. Indeed, experimental evidence was reported that the flux through a Sr$_2$RuO$_4$ ring is quantized in units of $\Phi_0/2$~\cite{jang:11}.
The rare observations of fractional flux quanta have been encountered only for unconventional superconductors. This is not an accident, because it is the existence of gap nodes, in particular for the $d$-wave pairing symmetry of the cuprates, which allows for the emergence of a PDW groundstate~\cite{loder:10}.

In the following we apply the concept of the generalized PDW with arbitrary combinations of COMMs $q_1$ and $q_2$ to flux-threaded hollow cylinders. We show that for sufficiently large cylinders, the groundstate energy $E(n\Phi_0/2)$ (with integer $n$) is minimized by a combination of two COMMs $q_1$ and $q_2$ fulfilling $(q_1+q_2)/2=n/2$, i.e., the flux is quantized in units of $\Phi_0/2$. The energy minima for different $n$ become degenerate in the thermodynamic limit, leading to a $\Phi_0/2$-periodicity of the groundstate energy and the supercurrent.

We start from a pairing Hamiltonian on a square lattice (the lattice constant is set to 1) with periodic boundary conditions, and $N_\theta$ sites in $\theta$-direction, $N_z$ sites in $z$-direction~\footnote{For simplicity, periodic boundary conditions are chosen in both $\theta$- and $z$-direction. In the context of flux quantization, the choice of boundary conditions in $z$-direction is irrelevant.}:
\begin{multline}
{\cal H}=\sum_{\k,s}\epsilon_\k(\varphi) c^\dag_{\k s} c_{\k s}\ +\\
\frac{1}{N_\theta N_z}\sum_\q\sum_{\k,\k'}\sum_{s,s'}V(\k,\k',\q)c^\dag_{\k s} c^\dag_{-\k+\q s'}c_{-\k'+\q s'}c_{\k' s}
\label{G1}
\end{multline}
with an attractive interaction $V(\k,\k',\q)$. Here, $\k=(k_\theta,k_z)$, where $k_\theta=1,\dots,N_\theta$ enumerates the angular momenta $\hbar k_\theta$ for motion around the cylinder, and $k_z=1,\dots,N_z$ the momenta $\hbar k_z$ in $z$-direction.
The single-electron dispersion takes the form
\begin{multline}
\epsilon_\k(\varphi)=-2t\,\left[\cos\left(\frac{2\pi (k_\theta-\varphi)}{N_\theta}\right)+\cos\left(\frac{2\pi k_z}{N_z}\right)\right]
\label{G2}
\end{multline}
with nearest-neighbor hopping amplitude $t$.
The magnetic flux enters in $\epsilon_\k(\varphi)$ through $\varphi=\Phi\,e/hc=\Phi/2\Phi_0$.

\begin{figure*}[t]	
\centering
\hspace{3mm}
\begin{overpic}
[width=1.95\columnwidth]{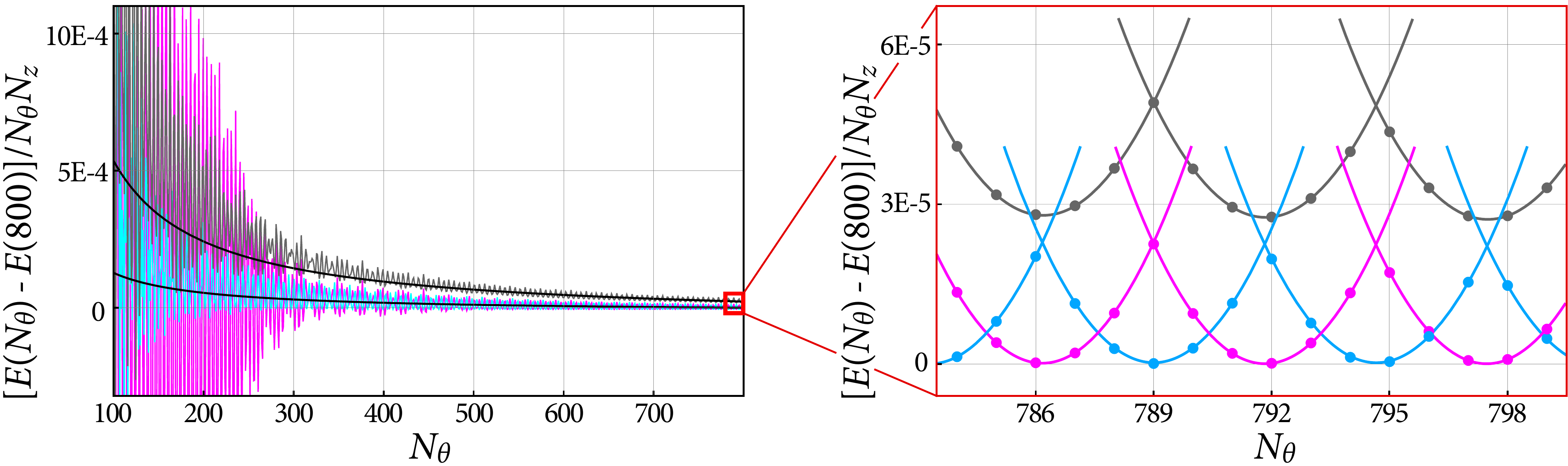}
\put(-5,59){(a)}
\put(-5,59){(b)}
\end{overpic}
\vspace{-1mm}
\caption{Energies $E_{q,-q+1}(1/4)/(N_\theta N_z)$ (blue), $E_{q,-q}(0)/(N_\theta N_z)$ (pink), and $E_{q,-q}(1/4)/(N_\theta N_z)$ (grey) per lattice site as a function of the circumference $N_\theta$ of the cylinder, for $N_z=80$, $V_1=2.2\,t$, $\rho=0.8$ and $t'=-0.3\,t$. Although $E_{q_1,q_2}(\varphi)$ oscillates with the cylinder's radius, for large enough cylinders $E_{q,-q}(0)=E_{q,-q+1}(1/4)$ is always smaller than $E_{q,-q}(1/4)$.\vspace{-3mm}}
\label{fig2}
\end{figure*}

For a superconducting state with singlet pairing, we use the BCS mean-field decoupling scheme and approximate $\langle c^\dag_{\k\ua}c^\dag_{-\k+\q\da}c_{-\k'+\q\da}c_{\k'\ua}\rangle\rightarrow\langle c^\dag_{\k\ua}c^\dag_{-\k+\q\da}\rangle\langle c_{-\k'+\q\da}c_{\k'\ua}\rangle$. The Heisenberg equations of motion for the spin independent imaginary time Green's function ${\cal G}(\k,\k',\tau)=-\langle T_\tau c_{\k s}(\tau)c^\dag_{\k's}(0)\rangle$ and the anomalous propagators ${\cal F}(\k,\k',\tau)=\langle T_\tau c_{\k s}(\tau)c_{-\k's'}(0)\rangle$ and ${\cal F}^*(\k,\k',\tau)=\langle T_\tau  c^\dag_{-\k s}(\tau)c^\dag_{\k's'}(0)\rangle$ for $s\neq s'$ then lead to the Gor'kov equations \cite{agd7}:
\begin{multline}
{\cal G}(\k,\k',\omega_n)={\cal G}_0(\k,\omega_n)\\\times\Bigg[\delta_{\k\k'}-\sum_\q\Delta(\k,\q){\cal F}^*(\k-\q,\k',\omega_n)\Bigg],
\label{G5}
\end{multline}
and
\begin{align}
{\cal F}(\k,\k',\omega_n)\!={\cal G}_0(\k,\omega_n)\!\sum_\q\Delta(\k,\q){\cal G}(-\k',-\k+\q,-\omega_n),
\label{G6}
\end{align}
where ${\cal G}_0(\k,\omega_n)=\left[i\omega_n-\epsilon_\k+\mu\right]^{-1}$ is the Green's function in the normal state and $\omega_n=(2n-1)\pi k_{\rm B}T$ is the fermionic Matsubara frequency. The average charge density $\rho$ is controlled by the chemical potential $\mu$. The order parameter $\Delta(\k,\q)$ represents electron pairs with center-of-mass momentum $\hbar\q$ and is determined self consistently as shown below.

At this point the form of the interaction is crucial. As the simplest ansatz that allows for unconventional pairing, we choose an attraction between electrons on neighboring sites.
With the restriction to singlet pairing only the extended $s$- and the $d$-wave channels remain, which leads to the interaction $V(\k,\k',\q)=V_+(\k,\k',\q)+V_-(\k,\k',\q)$ with 
\begin{align}
V_\pm(\k,\k',\q)=Vg_\pm(\k-\q/2)g_\pm(\k'-\q/2),
\label{G4}
\end{align}
where $g_\pm(\k)=\cos (2\pi k_\theta/N_\theta)\pm\cos (2\pi k_z/N_z)$. The OP thus becomes
\begin{align}
\Delta(\k,\q)=\Delta_s(\q)g_+(\k-\q/2)+\Delta_d(\q)g_-(\k-\q/2)
\label{G4.1}
\end{align}
with a non-vanishing extended $s$-wave contribution $\Delta_s(\q)$ if $\q\neq\bm0$~\cite{loder:10}.
Cooper pairs moving in the flux threaded cylinder acquire angular momentum and we therefore choose $\q=(q,0)$ along the $\theta$-direction. Inserting $V(\k,\k',\q)$ from (\ref{G4}) into~(\ref{G5}) generates a set of coupled self-consistency equations for $\Delta_s(\q)$ and $\Delta_d(\q)$:
\begin{align}
\Delta_{s,d}(\q)=-\frac{k_{\rm B}TV}{N_\theta N_z}\sum_{\k}g_{s,d}(\k-\q/2)\sum_n {\cal F}(\k-\q,\k,\omega_n).
\label{G13.0}
\end{align}

In an ansatz for a self-consistent solution, we choose $Q$ trial vectors $\q_1,\dots,\q_Q$ and set $\Delta(\k,\q)=0$ for all $\q\neq\q_i$. Thereby we test selected combinations of $\q$-vectors. This procedure and its solutions without magnetic flux are discussed in Ref.~[\onlinecite{loder:10}]. In the context of flux quantization, two notions are important: ($i$)~If instead of Eq.~(\ref{G4}) conventional $s$-wave pairing originating from an on-site interaction is considered, the groundstate is the standard BCS superconductor with one COMM and therefore no fractional flux quanta occur. ($ii$)~For the pairing interaction~(\ref{G4}), the groundstate for zero flux is either a $d$-wave BCS, or a PDW superconductor with the two COMMs $\q$ and $-\q$, i.e. $Q=2$, depending on the interaction strength $V$. If $V$ exceeds a critical value $V_{\rm c}$, a first order transition from the BCS to a PDW state occurs.
This phase transition originates from the competition between the cost in kinetic energy $E_\text{kin}$ and a gain in condensation energy $E_\text{con}$ arising from finite COMMs in the presence of gap nodes. The $d$-wave superconductor has four nodal points in $\k$-space where unpaired electrons remain at the Fermi energy. In the PDW state, these electrons can be paired by $\Delta(\k,\pm\q)$. The condensation energy is therefore optimized by the vector $\q$ for which the nodes in $\Delta(\k,\q)$ and $\Delta(\k,-\q)$ are furthest apart~\cite{loder:10}.

In the following we investigate the flux dependence of the generalized PDW and its total energy $E_{q_1,q_2}(\varphi)$ for the state with the two COMMs $\{q_1,q_2\}$ at $T=0$. In particular we search for the combinations $\{q_1,q_2\}$ for which $E_{\q_1.\q_2}(\varphi)$ is minimal for flux values $\varphi=n/2$, and we test whether combinations $\{\q_1,\q_2\}$ exist for which $E_{q_1,q_2}(\varphi)$ is minimal at $\varphi=n/4$. The charge density is fixed to $\rho=0.8$, and an additional next-nearest-neighbor hopping $t'=-0.3\,t$ is included. For these values and $\varphi=0$, a PDW state is realized for $V>V_{\rm c}\approx2.1\,t$ with $q_1=-q_2=q$ as the integer closest to $N_\theta/A$, where $A\approx6$. The value of $A$ depends on $\rho$ and typically shrinks with increasing interaction strength $V$~\cite{loder:10}. At the flux value $\varphi=1/4$ we expect a realization of COMMs with $(q_1+q_2)/2=2\varphi=1/2$. In particular, the combination $\{q,-q+1\}$ is a likely candidate for the groundstate. However, especially for small cylinders, the favored combination of COMMs may vary with system size; therefore different sets $\{q_1,q_2\}$ are separately compared for each system size.

Figure~\ref{fig2} shows the energy $E_{q,-q+1}(1/4)$ as a function of the circumference $N_\theta$ of the cylinder for $V_1=2.2\,t$ and, for comparison, $E_{q,-q}(0)$ and $E_{q,-q}(1/4)$. The groundstate COMM $q$ is the integer closest to $\sim N_\theta/A$ and switches to the next integer when $N_\theta$ increases by $A$. Since $A$ is typically not commensurate with the lattice, there is a sensitive size dependence for small cylinders ($N_\theta\lesssim300$) where the existence of an energy minimum at $\varphi=1/4$ is not certain. For large cylinders, the total energy oscillates in $N_\theta$ with periodic minima at $N_\theta=nA$ and $N_\theta=nA/2$. The amplitude of the oscillations decreases with $1/N_\theta$ and in the limit $N_\theta\rightarrow\infty$ one finds that $E_{q,-q}(0)=E_{q,-q+1}(1/4)$ is always smaller than $E_{q,-q}(1/4)$. 

\begin{figure}[t]	
\centering
\hspace{3mm}
\begin{overpic}
[width=0.97\columnwidth]{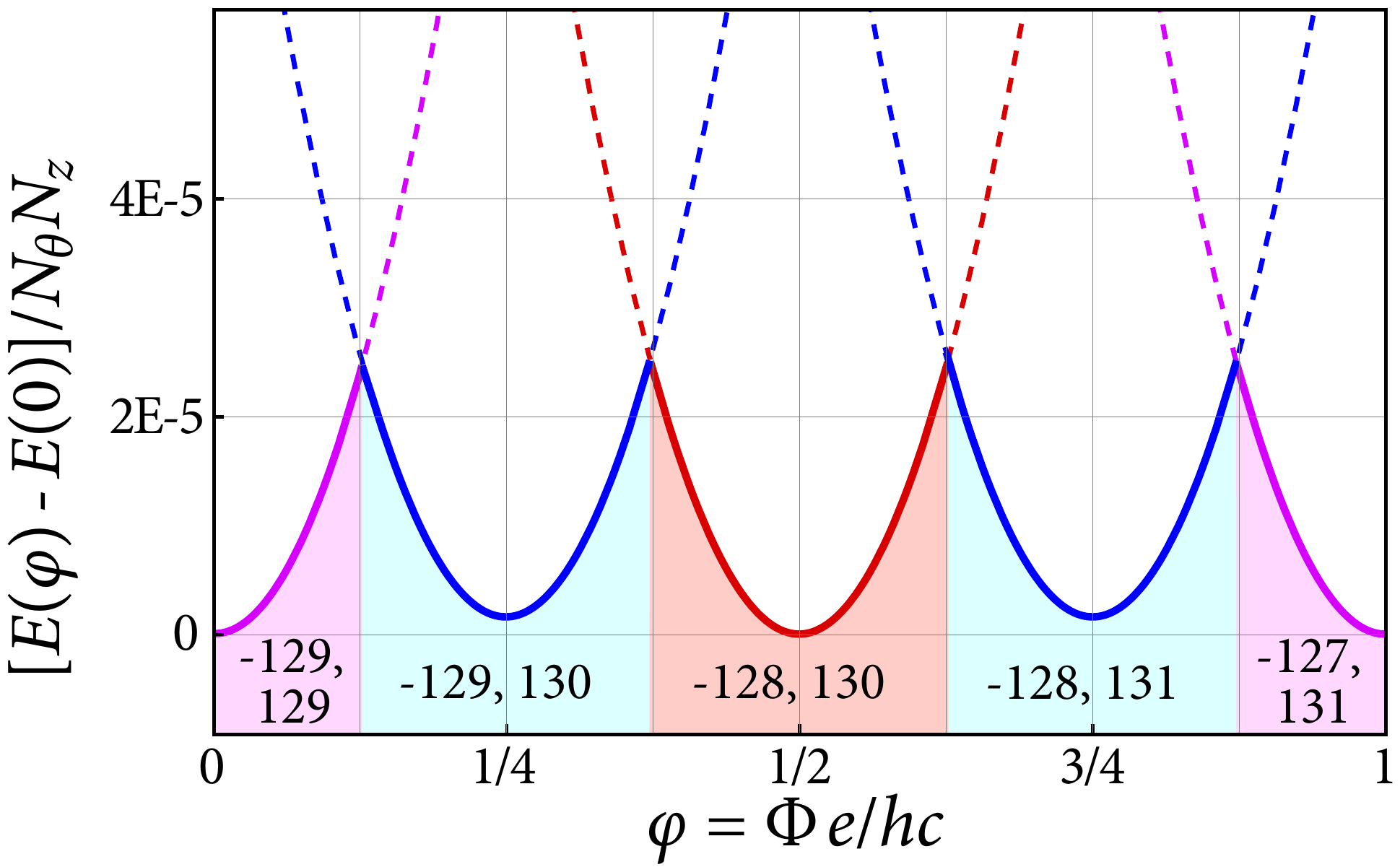}
\end{overpic}
\vspace{-1mm}
\caption{Groundstate energy $E(\varphi)=\min_{q_1,q_2}E_{q_1,q_2}(\varphi)$ for the same model parameters as in Fig.~\ref{fig2} with $N_\theta=796$. The numbers in the shaded areas indicate the COMMs $q_1$ and $q_2$ in the corresponding flux regimes. The energy difference between the blue and the pink minima corresponds to the difference between the blue and pink dots at $N_\theta=796$ in Fig.~\ref{fig2}. Wether the fluxless state is formed by two odd or two even COMMs depends on the system size $N_\theta$.\vspace{-3mm}}
\label{fig3}
\end{figure}

The SC groundstate around $\varphi=1/4$ is therefore realized for the combination  $\{q,-q+1\}$ and $E_{q,-q+1}(\varphi)$ is indeed minimal at $\varphi=1/4$. Upon approaching $\varphi=1/2$, the groundstate switches to the combination $\{q+1,-q+1\}$ with an energy minimum at $\varphi=1/2$, and to $\{q+1,-q+2\}$ around $\varphi=3/4$ (see~Fig.~\ref{fig3}). For an arbitrary flux value, the combination of COMMs minimizing the energy is given by
\begin{align}
q_1+q_2=\text{floor}\left(4\varphi+1/2\right),
\end{align}
where $\text{floor}(x)$ is the largest integer smaller than $x$.
The new energy minima correspond to fractional flux values $n\varphi/4$ and therefore lead to flux quantization in units of $\Phi_0/2$. This result remains valid also for thick-walled, irregular or disordered cylinders, which can be modeled by averaging the total energy over various channels with different circumferences. In the thermodynamic limit, the energy minima at flux values $\Phi=0$ and $\Phi=\Phi_0/2$ become degenerate, and consequently, the flux periodicity of thermodynamic properties like the supercurrent is $\Phi_0/2$ as well. This situation is illustrated in Fig.~\ref{fig3} for $N_\theta=796$, for which $E_{q,-q}(0)$ and $E_{q,-q+1}(1/4)$ are almost equal. The groundstate energy as a function of $\varphi$ forms a series of parabolae with minima at $\varphi=n/4$. The supercurrent is obtained as
\begin{align}
J(\varphi)=-\frac{e}{hc}\min_{q_1,q_2}\frac{\partial E_{q_1,q_2}(\varphi)}{\partial\varphi},
\label{G58}
\end{align}
and vanishes whenever the groundstate energy has a minimum. Therefore, the formation of a SC state with half-integer valued phase winding number $(q_1+q_2)/2$ compensates the current induced by the half-integer magnetic flux $\Phi=\Phi_0(q_1+q_2)/2$ and thereby minimizes the total energy of the system.   

In finite systems, SC states with even and odd phase-winding numbers are distinct and their energy minima at the corresponding flux values differ~\cite{loder:08}. The flux periodicity is therefore strictly $hc/e$, although quantization occurs in units of $hc/4e$.
We thus identify three distinct classes of pair states on a flux threaded cylinder (see Fig.~\ref{fig3}): superpositions of two odd (purple) or two even (red) COMMs with energy minima at integer multiples of $\Phi_0$, and even-odd (blue) combinations with energy minima at half-integer multiples of $\Phi_0$. These pair states correspond to integer or half-integer phase winding numbers of the SC order parameter.
In principle, states with more than two different COMMs are possible as well. Such states will allow for fractional flux quanta of magnitude $\Phi_0/Q$, but they are not realized as groundstates in the parameter range of our model Hamiltonian. In alternative Hamiltonians the prerequisite will remain the existence of an instability towards nodal superconductivity, which allows for fractional flux quantization due to coexisting COMMs.

In a real flux threaded SC ring, flux always penetrates into the superconductor itself. The system is then no longer periodic in $\Phi$ and the spatial variations of the magnetic field induce additional COMMs. Therefore, a unidirectional PDW state is expected to break down even in weak magnetic fields, leaving a standard $d$-wave superconducting groundstate. Above this threshold field, $\Phi_0/2$ flux-quantization will then disappear abruptly, returning to the flux quantum $\Phi_0$. As described above (c.f. Fig.~\ref{fig1}), in YBCO SQUIDs the periodicity of the supercurrent changes indeed abruptly at a threshold magnetic field~\cite{schneider:04}. The low-field $\Phi_0/2$ oscillations were also reported to vanish close to $T_{\rm c}$~\cite{schneider:04}. While the solutions of our model do not change qualitatively with temperature far below $T_c$, the PDW is also replaced by a $d$-wave BCS state upon approaching $T_{\rm c}$.

Whether a PDW state occurs in cuprate superconductors is unsettled. However, uniaxial charge and spin modulations in coexistence with superconductivity were verified  for 214 cuprates~\cite{tranquada:95,tranquada:08,hucker:11,sonier:07}, which inevitably imply spatial modulations in the pair density~\cite{raczkowski:07,yang:09,berg:09c,loder:11a,loder:11}.
In the typical rectangular geometry of the SQUID experiments, the orientation of an anticipated PDW would differ from the cylinder geometry discussed above. Although the emergence of SC states with half-integer phase-winding numbers and $\Phi_0/2$ periodicity will inevitably occur, too, the explicit solution of a model Hamiltonian in a SQUID geometry is more involved---especially for a finite mismatch angle between the lattices on both sides of the Josephson junctions. A possible analysis of this geometry will require numerical simulations.

\color{black}
\begin{acknowledgements}
The authors thank for a stimulating exchange of ideas with Jochen Mannhart and Christof Schneider. We gratefully acknowledge discussions with Daniel Agterberg and Raymond Fr\'esard. This work was supported by the DFG through TRR 80.
\end{acknowledgements}

\end{document}